# DETECTION OF A FAINT OPTICAL JET IN 3C 120[1]

Jens Hjorth[2,3], Marianne Vestergaard[4,5], Anton N. Sørensen[3,6], and Frank Grundahl[3]

## ABSTRACT

We report the detection of an optical jet in the nearby Seyfert 1 radio galaxy 3C 120. The optical jet coincides with the well-known radio jet and emits continuum radiation ($B$,$V'$,$I$) with a radio-to-optical spectral index of 0.65. There are no clear optical counterparts to the radio knots, although the optical condensation A of the galaxy, which includes the bright $4''$ radio knot, is found to be 12 % polarized with the electric field vectors perpendicular to the jet. These findings indicate that 3C 120 contains the 6th known extragalactic optical synchrotron jet, quite similar in its properties to the jet of PKS 0521−36. The outer parts of the jet is the faintest known optical jet and was discovered as the result of a dedicated effort to detect it. It is therefore possible that more optical jets can be discovered in systematic searches by combining deep imaging in the optical or near-IR with careful galaxy subtraction methods.

---

[1]Based on observations obtained with the *Nordic Optical Telescope* (*NOT*), La Palma.

[2]Institute of Astronomy, Madingley Road, Cambridge CB3 0HA, UK; jens@ast.cam.ac.uk.

[3]Institute of Physics and Astronomy, University of Aarhus, DK–8000 Århus C, Denmark; norup@obs.aau.dk, fgj@obs.aau.dk.

[4]Harvard–Smithsonian Center for Astrophysics, 60 Garden Street, MS #12, Cambridge, MA 02138, USA; vester@head-cfa.harvard.edu.

[5]Copenhagen University Observatory, Øster Voldgade, DK–1350 Copenhagen, Denmark.

[6]Nordic Optical Telescope, Observatorio Roque de los Muchachos, Apartado 474, E–38700 Santa Cruz de La Palma, Spain.

## 1. INTRODUCTION

Extragalactic optical jets are counterparts to radio jets resulting from energy transport in collimated beams out of the nuclei of powerful radio galaxies. The polarized nature of the jets is best explained by synchrotron radiation from an accelerated electron plasma. The existence of optical synchrotron radiation implies that the electrons responsible for the emission must be reaccelerated inside the jet because their life-time is much too short for radiation to be detected otherwise. Optical jets thus offer the prospect of identifying the locations within the jet at which relativistic particles are generated or reaccelerated. Moreover, optical emission is free from Faraday rotation effects which means that an ambiguity in magnetic field mapping is removed. The energy loss time scales for X-ray emitting electrons are even shorter and so, X-ray jets are more constraining to the need for local acceleration of the electrons. The big advantage in the optical, however, is the possibility of doing polarization measurements, high-resolution imaging, and spectroscopy.

The maximum energy that an electron can reach in the reacceleration process corresponds to a cut-off frequency. Together with the apparent luminosity of the jet and the spatial resolution attained in the observations, this parameter determines whether a jet is visible at optical wavelengths. Only few optical counterparts to radio jets are known (M87, 3C 273, 3C 66B, PKS 0521−36, 3C 264, see Crane et al. 1993), and their optical spectra show the presence of a cut-off at about $10^{15}$ Hz, i.e., in the optical region (Keel 1988). The radio spectral indices do not vary much amongst radio jets and are in general close to $\alpha_{\rm r} \equiv \alpha_{\rm radio-radio} \approx 0.6$ [$F_\nu \propto \nu^{-\alpha}$, i.e., $\alpha_{1-2} \equiv -\log(F_2/F_1)/\log(\nu_2/\nu_1)$], but the cut-off frequency makes the radio-to-optical spectral index ($\alpha_{\rm ro}$) vary from one jet to another. In optical jets one has $\alpha_{\rm ro} \approx 0.7 - 0.9$. The very small number of optical counterparts to radio jets may be due to the cut-off frequency being in general $\lesssim 10^{14}$ Hz. The question that arises is whether the few known optical jets belong to the tail of the distribution of cut-off frequencies among radio jets, or if they are well separated from this distribution. If the first hypothesis is correct, then *more optical jets should be detectable at fainter levels* (Fraix-Burnet et al. 1991b).

To look into this question we observed the nearby ($z = 0.033$) Seyfert 1 radio galaxy 3C 120 (II Zw 14, PKS 0430+05). 3C 120 is an enigmatic galaxy which has been extensively studied for its many different characteristics. The galaxy harbors a variable AGN of moderate luminosity and is a powerful X-ray source (e.g., Maraschi et al. 1991). It has an interesting radio structure (e.g., Walker, Walker, & Benson 1988): A jet showing superluminal motion is seen on scales ranging (in projection) from 0.5 pc to 400 kpc with the direction of propagation of the jet rotating by nearly 180°. Its average spectral index is $\alpha_{\rm r} = 0.65$ (Walker, Benson, & Unwin 1987). The underlying galaxy is rich in extended emission condensations, some possibly ionized by the non-thermal UV continuum flux from the nucleus, while others could be gas photoionized by hot, newly produced stars—some perhaps induced by precession of the jet (Soubeyran et al. 1989).

The galaxy 3C 120 has acted as a prime target in the search for optical jets in powerful radio galaxies (e.g., Fraix-Burnet et al. 1991b; Fraix-Burnet, Golombek, & Macchetto 1991a), but previous attempts to detect an optical synchrotron counterpart to the radio jet in this object have been plagued by the above mentioned condensations and have been limited by the attained spatial resolution or detection level. An optical emission-line signature of the jet was seen by Axon et al. (1989) who found evidence for a direct interaction of the jet with the interstellar medium. However, the search for an optical synchrotron jet has been unsuccessful insofar as no pure optical continuum emission at the exact location of the radio knots has been detected (but see Lelièvre et al. 1994), leading Fraix-Burnet et al. (1991a) to conclude that there is probably no optical counterpart to the radio jet. Through deep high-resolution imaging the observations presented in this *Letter* were designed in such a way that a non-detection of the jet would rule out its existence if it had a normal spectral index. That is, with $\alpha_{\rm ro} < 0.9$ the jet would be detected in our observations.

## 2. OBSERVATIONS AND IMAGE ANALYSIS

A deep high spatial resolution imaging study was carried out in two broad bands essentially free of line-emission ($B$ and $I$). Using the 2.56-m *Nordic Optical Telescope* (*NOT*) we imaged 3C 120 during half a night on Oct. 17 1993 with the IAC CCD camera equipped with a Thomson $1024 \times 1024$ chip giving a pixel scale of $0\farcs139$ pixel$^{-1}$ and a read-out noise of 6 e$^{-}$. Several shifted images in each band were secured in order not to saturate the bright ($B = 14.8$, $I = 13.4$) core, with typical exposure times in $B$ and $I$ of 1200 s and 180 s, respectively. A total exposure time of $\sim$5500 s was obtained in each band. The final bias corrected, flat fielded, cosmic-ray cleaned, coadded images have resolutions of $0\farcs84$ in $B$ and $0\farcs76$ in $I$ (FWHM).

The photometric calibration was performed using the internal calibration stars (1–4) present in the exposed field of the galaxy (Moles et al. 1988). No color term was assumed (Aparicio & Gallart 1993) as the filters used are close to the standard Johnson system. The zero point of the calibration is accurate to $\pm 0.05$ mag. Correction for Galactic and intrinsic extinction was performed following Moles et al. (1988). Adopting $E(B-V)$ of 0.20 mag, this is equivalent to $E(B-I) = 0.50$ mag, assuming that the extinction internal to 3C 120 follows the standard interstellar reddening law (e.g., Zombeck 1990). Evidently, the intrinsic extinction of 3C 120 [$E(B-V) = 0.05$] is uncertain so, unless otherwise noted, apparent optical magnitudes presented in this *Letter* are uncorrected for this extinction (with corrected values given in parentheses), whereas all intrinsic properties of the jet such as optical fluxes and spectral indices are corrected for extinction.

In order to detect faint features in the galaxy a smooth fit to the underlying galaxy was subtracted. Careful galaxy subtraction is crucial since artifacts may otherwise easily show up. We used a dedicated galaxy fitting routine developed by B. Thomsen (Thomsen & Baum 1989; Sodemann & Thomsen 1994) to obtain a smooth fit of a galaxy rather than a fit to isophotes. It calculates radial harmonic profiles by means of a robust regression technique that fits Fourier series to circular or elliptical rings centered on the object and allows high order fits as well as masking of knots, jets etc. In the construction of the galaxy fit the very bright core was first subtracted using DAOPHOT II and ALLSTAR (Stetson 1987). Then the galaxy fit was established iteratively: A low order fit was subtracted from the image, the most conspicuous knots were masked out, and a new and better fit was computed. When repeating this procedure more and more features gradually showed up in the residuals as new fits were subtracted from the galaxy. Those which appeared robust were masked out and higher order fits were performed. After about 5 iterations the fit was stable and the residuals were robust to changes in the fitting parameters. We have confidence in our fits as our independent fits in $B$, $V'$, and $I$ yield very similar residuals with almost no signs of 'ringings', i.e., there are practically no regions of negative intensity.

The final residuals of the $I$ image with the core added back are shown as a contour plot in Figure 1 with a 4885 MHz radio overlay. The residual images reproduce all previously known condensations (labeled in Fig. 1 as in Fraix-Burnet et al. 1991a) plus fainter signatures, in particular a jet-like feature extending from $6''$ to $15''$ from the core (labeled 'J') at the exact location of the radio jet and with the same bend. [Note that the strong emission to the north-west (condensations A, B, and D) should not be mistaken for the optical jet (Arp 1981; Wlerick et al. 1981).] The present *Letter* concentrates on this jet, 'J'; the properties of the rest of the galaxy and its condensations are discussed in a separate paper (Vestergaard, Sørensen, & Hjorth 1995).

Observations for imaging polarimetry were obtained at the *NOT* during December 1993, in grey-time under non-photometric conditions with average seeing of $1\farcs2$. The Stockholm CCD camera with a $512 \times 512$ chip, giving a pixel scale of $0\farcs2$ pixel$^{-1}$ and a read-out noise of 20 e$^{-}$, was used. The exposure time totals 36 900 s through a linear polarizer (at four different orientations) and a $V'$ filter. The $V'$ filter transmits more than half the peak transmission between 530 nm and 600 nm, transmits less than 5 % of the peak transmission at the redshifted [OIII]λ5007 line, and thus mainly transmits continuum radiation. Exposures through the $V$ and $V'$ filters without the polarizer were also obtained. As the combined $V'$ image is of lower resolution ($0\farcs92$) and sensitivity than the $B$ and $I$ images and is difficult to calibrate, it is only used for the construction of a color image in this *Letter*.

For calibration, the star HD 25104 and the reflection nebula NGC 2261 were compared to values from Turnshek et al. (1990). To compensate for non-photometric conditions, the star $20''$ E of the nucleus and a smooth galaxy-fit were used as zero points for the polarization. In the galaxy-subtracted residual images the polarizations of the detected features were measured. The uncertainties were calculated from photon statistics and were increased by a factor estimated from difficulties in determining the background level. Details of the polarization measurements are given in Vestergaard et al. (1995).

## 3. RESULTS

An [O III] image constructed from the $V$ and $V'$ images (Vestergaard et al. 1995) shows good agreement with the direct narrow-band [O III] image of Hua (1988) which reveals no jet-like features in the region of 'J' (Fig. 1). The optical emission-line radiation coincident with the radio jet extends to about 6–7$''$ from the nucleus whereas the optical continuum radiation under discussion here is located farther from the core. The emission-line counter jet (Axon et al. 1989) is not seen either, and the strong H II regions (Hansen, Nørgaard-Nielsen, & Jørgensen 1987) are very faint. This, together with the fact that the independent residuals in the $B$, $V'$, and $I$ are very similar, confirms that the wavelength regions covered by the filters used are almost uncontaminated with emission lines at $z = 0.033$, i.e., the images definitely show the continuum emitting features in the galaxy.

A blow-up of the jet region of Figure 1 is shown as a composite color image in Figure 2. It is clear that the gross structure of the optical jet coincides with that of the radio jet, although there are no clear optical counterparts to the radio knots in this image. An exception may be the radio knot $4''$ W of the nucleus which is located on top of the southern part of condensation A. This part of the condensation is detected to be significantly redder than its immediate surroundings (by $\sim 0.8$ mag in $B - I$).

The optical jet ('J') that we have discovered extends out to $15''$ corresponding to a projected distance of 9 kpc from the nucleus (for $H_0 = 75$ km s$^{-1}$ Mpc$^{-1}$). Thus, there is no optical counterpart to the $20''$ radio knot, presumably because it may have a smaller cut-off frequency or a steeper spectral index, an effect also seen in the outer parts of M87 (Zeilinger, Møller, & Stiavelli 1993). The average spectral indices for the optical jet (computed in apertures 4.″6 wide, between 5.″6 and 15″ from the core, i.e., excluding the 4″ radio knot and condensation A) are $\alpha_{\rm ro} \equiv \alpha_{I-{\rm radio}} = 0.65$, $\alpha_{B-{\rm radio}} = 0.69$, and $\alpha_{\rm opt} \equiv \alpha_{B-I} = 1.3$ (the corresponding values uncorrected for extinction are $\alpha_{I-{\rm radio}} = 0.69$, $\alpha_{B-{\rm radio}} = 0.73$, and $\alpha_{B-I} = 2.0$). The jet is very faint. The extinction-corrected integrated flux in the $B$ band is 14 $\mu$Jy, and the uncorrected value (6 $\mu$Jy) is slightly smaller than the previously faintest known jet in 3C 66B (Crane et al. 1993). The surface brightness is 26.2 (25.4) mag arcsec$^{-2}$ corresponding to an extinction-corrected flux of 0.27 $\mu$Jy arcsec$^{-2}$. The total magnitudes of the jet are $B = 22.0$ (21.1) and $I = 20.0$ (19.6), i.e., the jet is quite red, $B - I = 2.0$ (1.5). The characteristic parameters for the optical jet are summarized in Table 1.

The polarization in a 4″ × 2″ aperture oriented roughly N–S covering condensation A is 12% ± 4%, and the position angle (PA) of the electric field vector is 178° ± 6°. Also, a 2″ × 2″ field 6″ W of the nucleus (between condensation A and the optical jet 'J') is measured to have a polarization of 100% ± 30% at PA 0° ± 15°. The polarization of 'J', measured in a 8″ × 3″ box along the jet, centered 11″ W of the nucleus, is 15% ± 15%, with PA = 155° ± 20°. A subdivision into two apertures yielded consistent measurements for both. For such low significance levels, the polarization bias calculated by Wardle & Kronberg (1974) becomes relevant. The corrected value is 0% ± 15%, i.e., an upper limit of 15% is set on the polarization at the location of the optical jet.

## 4. DISCUSSION AND CONCLUSION

The detection of polarization in condensation A (of which the southern part coincides with the 4″ radio knot) links the optical features that we see to the radio jet. The direction and magnitude of the polarization is consistent with the findings of Walker et al. (1987) at 5 GHz. This indicates that it may be optical synchrotron radiation. However, as the polarization is perpendicular to the jet as well as to the line of sight to the nucleus, it may also be due to scattered light from the nucleus.

In contrast to the jet of M87, Walker et al. (1987) find that the Faraday rotation and depolarization of the 3C 120 jet are very small. This allows for a direct comparison between the radio polarization maps and the optical data. Walker et al. (1987) find some structure in the 4″ knot that may be associated with condensation A, with polarization ranging from almost zero to 15%. In this knot, and along the rest of the jet, the polarization is roughly perpendicular to the jet, without the strong local deviations found in the M87 jet (Fraix-Burnet, Le Borgne, & Nieto 1989). As a consequence, a strong increase in measured polarization with higher spatial resolution (as in M87) due to the small-scale structure in the polarization is unlikely to occur in 3C 120. The blue color of condensation

A (Fraix-Burnet et al. 1991a; Vestergaard et al. 1995) indicates that the majority of the light is from young stars, or is scattered light from the nucleus. Star light would tend to 'dilute' the measured optical polarization of condensation A to somewhat less than the maximum 15% measured in the radio. This may indicate that the measured polarization is better explained by scattered light from the nucleus. The facts that (i) condensation A turns below the radio-jet in our high-resolution images, i.e., it does not follow the radio jet into the core, and (ii) there is no indication of polarization gradients within the box used to measure the polarization, support this interpretation.

Such a putative reflection region may extend further away from the nucleus than condensation A and be the source of the strong polarization measured between the knot and the optical jet. The measured polarization of $100\% \pm 30\%$ is much larger than any radio polarization measured by Walker et al. (1987), but is consistent with that expected from an optically thin reflection nebula illuminated by the nucleus. An intriguing possibility is that the apparent coincidence of this region with strong [O III] emission (Hua 1988) marks the point where interaction between the jet and the ISM triggers the optical jet.

Lelièvre et al. (1994) measure continuum emission ($\sim V'$) and optical knots along the jet at distances between $2''$ and $7''$ from the nucleus and interpret this as optical jet emission. In our data we are unable to discern the light from a possible jet from that of condensation A and its extension into the core. In particular, our data provide no evidence for the presence of optical knots, but rather a smooth intensity distribution around condensation A (see Fig. 2). For example, Lelièvre et al. (1994) find a knot $6\overset{''}{.}66$ west of the core with a flux of 2.9 $\mu$Jy (in $1\overset{''}{.}25 \times 1\overset{''}{.}25$), while we measure fluxes of $0\overset{''}{.}29$ $\mu$Jy arcsec$^{-2}$ in $B$ and $0\overset{''}{.}66$ $\mu$Jy arcsec$^{-2}$ in $I$.

Outside condensation A the new optical jet in 3C 120 is safely detected. Its spectral indices are consistent with what is expected from synchrotron radiation [see the comparison of the 5 previously known extragalactic optical jets of Crane et al. (1993)] and larger than what is expected if the jet is the result of light from the nucleus scattered by dust. In fact, the deduction that the radio spectral index and the extinction-corrected radio-to-optical ($I$) spectral index have identical values (0.65) lends strong support to this interpretation. The fact that the counter jet is seen in the gas (Axon et al. 1989) but not in our continuum images is an extra argument for the continuum jet being (beamed) synchrotron radiation.

Concerning the morphology of the jet, the radio knots do not seem to have any clear optical counterparts as in e.g. M87, except perhaps at the $4''$ knot (this also means that we cannot decide whether the jet is resolved in our observations). A similar effect is seen in 3C 273 where the morphology of the jet is quite different at optical and radio wavelengths (Thomson, Mackay, & Wright 1993). It is possible that the jet of 3C 273 is not a pure synchrotron jet, and the same may be true for 3C 120. On the other hand, it is striking to what extent the parameters of the jet in 3C 120 (especially those uncorrected for extinction, see §3) resemble those of the jet of PKS 0521−36 (see Table 1 of Crane 1993), except for the larger surface brightness of the latter. Morphologically, PKS 0521−36 also has a smooth optical appearance and has been described as an 'aging' counterpart to M87 (Keel 1986). A detailed comparison of the radio and optical structures of this system only became possible with *HST* observations (Macchetto et al. 1991). Thus, a full discussion of the morphology (and, hence, the nature) of the optical jet in 3C 120 will have to await *HST* observations.

The jet is the faintest of the 6 known optical jets and resides in a Seyfert 1 galaxy (while other jets are harbored by ellipticals). This opens up the possibility that there may be many more optical jets waiting to be discovered and supports the view (e.g., Keel 1988; Fraix-Burnet et al. 1991b) that the visibility of the jet in the optical (besides apparent luminosity) sensitively depends on the exact location of the cut-off frequency. Indeed, many objects could have cut-off frequencies in the infrared (IR).

The jet of 3C 120 roughly has the same average color as the background sky (but is 4 magnitudes fainter—the galaxy does not contribute significantly to the background level outside 4″). Thus $S_I/S_B \sim B_I/B_B \sim N_I^2/N_B^2$ and hence $(S/N)_I \approx \sqrt{B_I/B_B}(S/N)_B$ ($S$=signal in jet, $B$=background level, $N$=noise) which shows that the jet is much easier detected in $I$ (this is due to the cut-off frequency). Furthermore, the jet is more easily discerned from the other condensations in $I$ than in $B$. In this *Letter* we have shown that even with a very complicated galaxy morphology it is possible to perform a reliable galaxy subtraction which nicely reveals the jet. Thus, a strategy to search at red optical or near-IR wavelengths, where the jets are brightest, could reveal several new jets. Candidates for such a study are low-redshift objects with small spectral indices and strong radio jet emission. A larger sample of extragalactic optical synchrotron jets could further constrain models for electron acceleration and energy emission, transport, and loss to help us understand better these interesting phenomena. Unfortunately, it will be hard to establish the synchrotron nature of very faint jets since polarization is difficult to measure at such low intensities.

In conclusion, we have discovered an optical counterpart to the radio jet outside 6″ from the core of the radio galaxy 3C 120 ($z = 0.033$). The jet emits continuum radiation and is detected in the $B$, $V'$, and $I$ bands. The optical-to-radio spectral index is 0.65, consistent with synchrotron radiation as in the 5 previously known extragalactic optical jets. Regardless of the nature of condensation A the detection of 12 % polarization is strong evidence that it is related to the radio jet. The higher spatial resolution and lower background level attainable with *HST* is crucial for future studies of the morphology and polarization of condensation A, the intermediate high-polarization region, and the optical jet. Also, the jet should be detectable with *ROSAT* HRI if the optical-to-X-ray spectral index is less than 1.3.

We thank Peter Barthel, Mark Birkinshaw, Bob Fosbury, Leif Hansen, Dan Harris, Gérard Lelièvre, Bjarne Thomsen, Bob Thomson, and Diana Worrall for helpful comments and discussions, Craig Walker for making the radio image electronically available, and Ramón García López and IAC for use of the IAC CCD camera prior to commissioning. We acknowledge support from the Carlsberg Foundation (JH) and from the Danish Natural Science Research Council (JH, MV).

Table 1. Characteristic Parameters for the Optical Jet in 3C 120

| Redshift | distance[a] (Mpc) | size (kpc) | $\alpha_{\rm opt}$ | $\alpha_{\rm ro}$ | $\alpha_{\rm r}$ | $F_\nu$ (6 cm) (Jy) | $F_\nu$ ($B$) ($\mu$Jy) | $F_\nu$ ($I$) ($\mu$Jy) |
|---|---|---|---|---|---|---|---|---|
| 0.033 | 133 | 9.1 | 1.3 | 0.65 | 0.65[b] | 0.05[c] | 14[d] | 34[d] |

[a]For $H_0 = 75$ km s$^{-1}$ Mpc$^{-1}$ and $q_0 = 0.5$.

[b]Walker et al. (1987).

[c]This corresponds to 0.02 Jy at 2 cm assuming $\alpha_{\rm r} = 0.65$.

[d]Corrected for internal and Galactic extinction, $E(B - I) = 0.50$ mag.

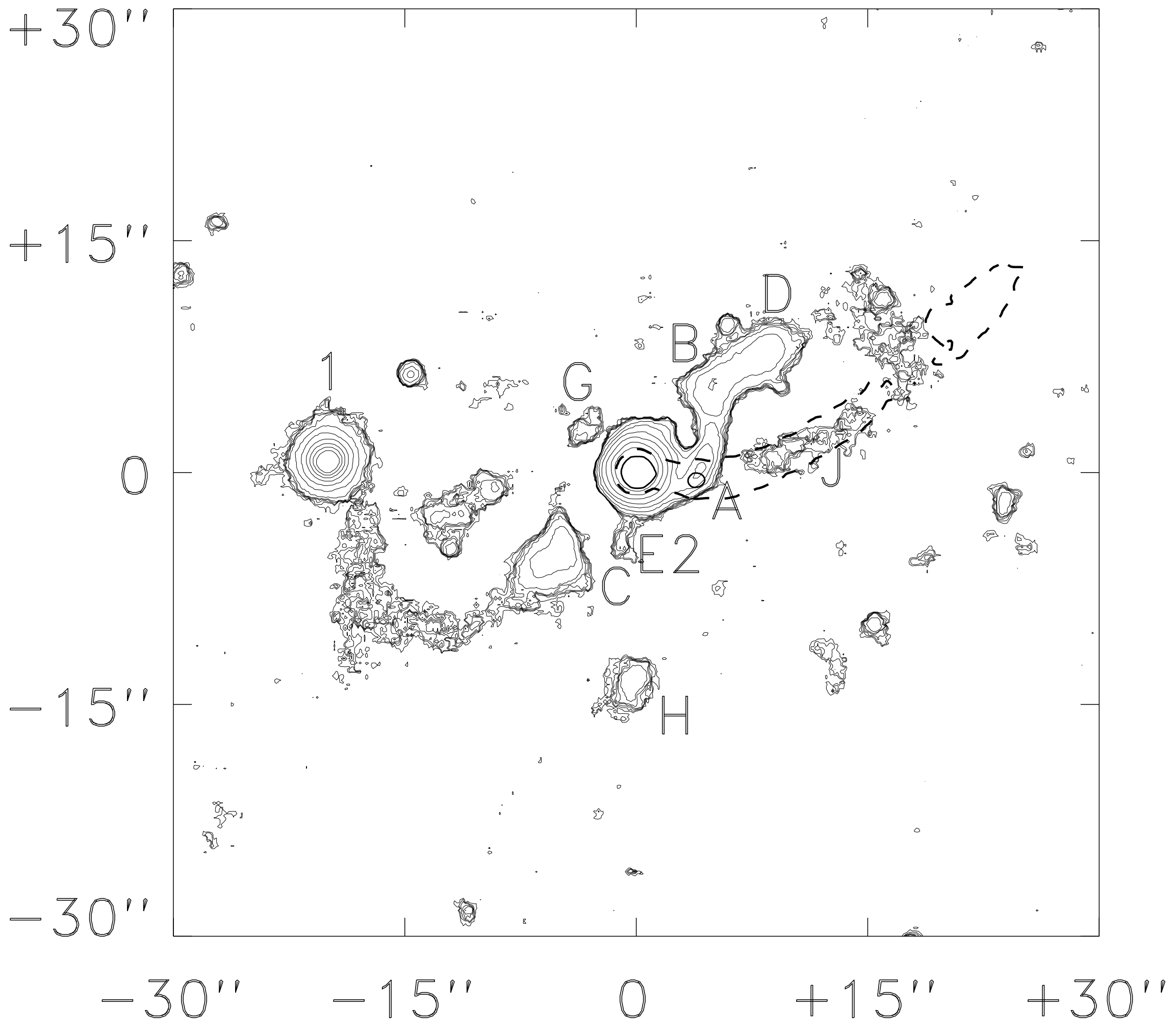

Fig. 1.— This contour plot shows the residuals of the $I$ image after the subtraction of the galaxy and taking the median in a $5 \times 5$ pixels region around each pixel. The contour levels are 0.7, 0.9, 1.1, 1.3, 1.5, 2, 4, 8, 16, 32, 64, and 128 $\mu$Jy arcsec$^{-2}$. The overlay (thick dashed curve) shows the outer contour (0.4 mJy arcsec$^{-2}$) and the 4″ knot (40 mJy arcsec$^{-2}$) of the 4885 MHz radio map shown in Fig. 3 of Walker et al. (1988), convolved to a resolution of ∼0.″8 FWHM to match the optical image. North is up and east is to the left. All the features seen are real (i.e., they are not artifacts of the galaxy subtraction); the brightest ones have been labeled and discussed by Soubeyran et al. (1989) and Fraix-Burnet et al. (1991a). A full discussion of the condensations is presented in Vestergaard et al. (1995).

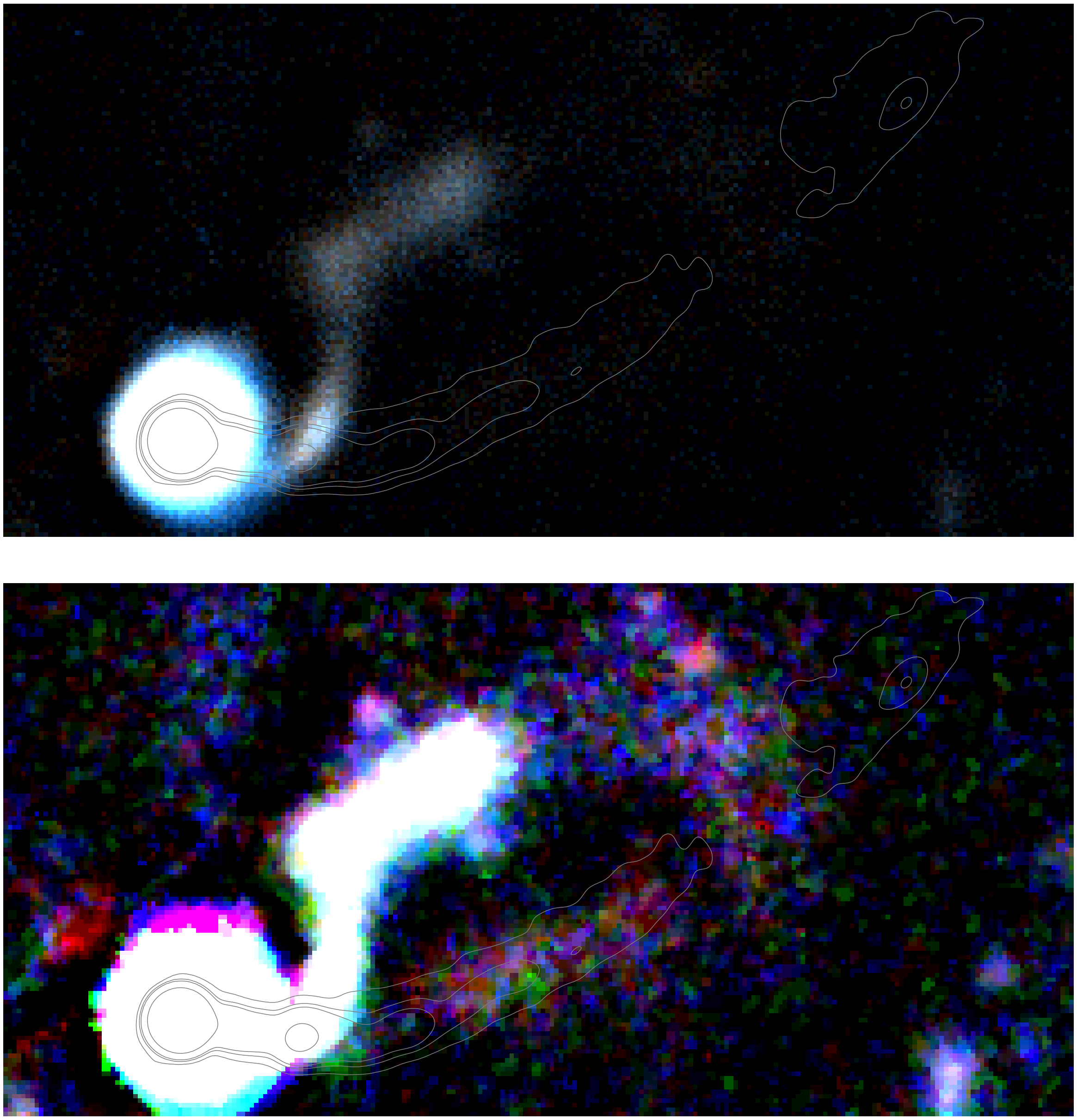

Fig. 2.— This figure shows the optical jet compared to the radio jet. The top panel is a composite color image made from the $B$ and $I$ images. It shows the core, condensation A and its eastern extension turning below the core, and the north-west extension into condensations B and D (see Fig. 1). The 4885 Mhz radio overlay reveals the coincidence of the 4″ knot with the southern part of condensation A. The resolution is $\sim 0\farcs8$ FWHM. The bottom panel displays the faint features of the jet region in more detail. This image is produced using the $B$, $V'$, and $I$ images smoothed to a resolution of $\sim 1''$ FWHM to reduce the noise. The colors of the outer parts of very bright regions may be misleading due to the images having slightly different PSFs. In both panels north is up and east is to the left and the images are $33\farcs3\times16\farcs6$.